\documentclass[12pt]{iopart}

\usepackage{xcolor}
\begin{document}
	
	\title{Mass spectra of $\Xi_{cc}$, $\Xi_{bc}$, $\Omega_{cc}$, and $\Omega_{bc}$ baryons in Regge Phenomenology}
	
	\author{Juhi Oudichhya, Keval Gandhi, and Ajay Kumar Rai}
	
	\address{Department of Physics, Sardar Vallabhbhai National Institute of Technology, Surat, Gujarat-395007, India}
	\ead{oudichhyajuhi@gmail.com}
	\vspace{10pt}

	\begin{abstract}
		In this article, we study the mass spectra of baryons containing two heavy quarks; charm-charm ($cc$) and bottom-charm ($bc$) with a light quark ($u,d,s$) within the framework of Regge phenomenology. With the assumption of linear Regge trajectories we have derived the relations between slope ratios, intercepts, and baryon masses. Using these relations, the ground state masses of  $\Xi_{cc}$, $\Xi_{bc}$, $\Omega_{cc}$, and $\Omega_{bc}$ baryons are obtained. The values of Regge slopes and Regge intercepts are extracted for these baryons to estimate the excited state masses in both the ($J,M^{2}$) and ($n,M^{2}$) planes. Our obtained results are compared with the experimental observations where available and other theoretical predictions, which could be a valuable addition to the interpretations of experimentally unknown heavy baryon spectra.
	\end{abstract}
	%
	\vspace{2pc}
	\noindent{\it Keywords}: Regge Phenomenology, hadron spectroscopy, doubly heavy baryons
	%
	%
	%
	%

	\section{Introduction}
	
	Due to the large number of data recently reported by many world-wide experimental facilities such as LHCb, CMS, SELEX etc., the study of baryons containing heavy quarks ($c,b$) is of significant interest. Many singly heavy baryons have been discovered in experiments so far and the quantum numbers of the observed states have been assigned \cite{PDG}. Also, two heavy quarks ($b$ or $c$) and a light quark ($u$, $d$, or $s$) can be bound together to form doubly heavy baryons in the constituent quark model, have two families: $\Xi$ and $\Omega$. $\Omega$  possesses a light strange quark whereas $\Xi$ has up or down quark(s), with two heavy quarks ($c$ and $b$). The study of these particles is crucial to understanding the hadron spectroscopy and Quantum Chromodynamics (QCD) at low energies.  In experiments, the first doubly heavy baryon was discovered with the charm quantum number $C=2$ is $\Xi_{cc}^{++}$ ($ucc$) \cite{LHCbCascc}. Although the SELEX Collaboration \cite{SELEXCas_cc1,SELEXCas_cc2} reported the observation of $\Xi_{cc}^{+}$ ($dcc$) at a mass of 3519$\pm$2 MeV years ago, other Collaborators such as FOCUS \cite{FOCUS}, BaBar \cite{BaBar}, and Belle \cite{Belle} failed to observe a state with the same properties. Recently in 2017, the LHCb Collaboration confirmed the existence of the $\Xi_{cc}^{++}$ baryon with a mass of 3621.40 $\pm$ 0.72(sat.) $\pm$ 0.27(syst.) $\pm$ 0.14($\Lambda_{c}^{+}$) through the decay $\Xi_{cc}^{++} \rightarrow \Lambda_{c}^{+}K^{-}\pi^{+}\pi^{+}$ \cite{LHCbCascc} and later this state was confirmed in the decay to $\Xi_{c}^{+}\pi^{+}$ \cite{LHCBCas2018}. Later, its lifetime time, mass, and production cross section were measured \cite{6,7,8}. Other than doubly charmed baryons, a different kind of doubly heavy baryon, the beauty-charmed baryon containing one bottom quark ($b$) and one charm quark ($c$) should also exist. The bottom-charm system is expected to be smaller than the charm-charm system and behave more like a point-like particle compared to the charm-charm system. Furthermore, beauty-charmed baryons incorporate more energy scales, including the bottom mass, charm mass, and the nonperturbative QCD scale $\Lambda_{QCD}$. Because of their significance, experimentalists have been focusing on searching for the beauty-charmed baryons. In 2020, the LHCb Collaboration seek for the doubly heavy $\Xi_{bc}^{0}$ baryon using its decay to the $D^{0}pK^{-}$ final state is performed using proton-proton collision data between 2016 and 2018. No significant signal is found in the invariant mass range 6.7 to 7.2 GeV/$c^{2}$ \cite{LHCbcas_bc}. Recently, in 2021 the LHCb Collaboration reported the first search for the $\Omega_{bc}^{0}$ baryon and a new search for the $\Xi_{bc}^{0}$ baryon in the mass range from 6.7 to 7.3 GeV/$c^{2}$, using \textit{pp} collision. No evidence is observed in this mass range \cite{Omega_bc2021}.

	In last few years researcher's have been putting their efforts to study the properties of  doubly heavy $\Xi$ and $\Omega$ baryons using various phenomenological and theoretical models \cite{ShiYJ2020,Qin2022,Mathur N2018,Azizi2019,Salehi2018,Nasrin2018}. The authors of Refs. \cite{Zalakcas_cc,ZalakOmega_cc} employed the hypercentral constituent quark model with Coulomb plus linear potential to obtain the mass spectra for doubly heavy $\Xi$ and $\Omega $ baryons. In this work, the first order correction is also taken into account to the potential. The ground, radial, and orbital states are calculated and the study of the Regge trajectories is performed in the $(J,M^{2})$ and $(n,M^{2})$ planes. Recently, these authors revisited the mass spectra of $\Xi_{cc}^{++}$ and $\Xi_{cc}^{+}$ baryons using Hypercentral Constituent  Quark Model  with Coulomb plus screened potential and also they calculate the
	transition magnetic moments of all the doubly heavy $\Xi$ and $\Omega$ baryons \cite{Zalak_universe2021}. Q. -X. Yu \textit{et al.} study the masses of the doubly heavy baryons within the framework of Bethe-Salpeter (BS) equations and solved them numerically with the Kernel containing the scalar confinement  \cite{Q.-X. Yu2019}. The authors of Ref. \cite{Yoshida2015} used the constituent quark model to study the singly and doubly heavy baryons with two exceptions; a color Coulomb term depending on quark masses and an antisymmetric L.S force. They studied the low-lying negative parity states and structures. The Ref. \cite{S. Gershteincas,S. Gershteinomega} obtained the mass spectra for the families of doubly heavy baryons $\Xi_{QQ'}$ and $\Omega_{QQ'}$ in the framework of the non-relativistic quark model with the QCD potential. They assume the quark-diquark structure for the wave functions and also the spin dependent splittings are taken into account.
	
	Another approach of QCD sum rules is used to obtain the masses and residues of radially and orbitally excited doubly heavy baryons \cite{Aliev2019}. The authors of Ref \cite{Ebert2002} used the relativistic quark model to  calculate  the mass spectra of baryons made up  of two heavy and one light quark. They assumed the light-quark–heavy-diquark structure of the baryon and under this assumption, the ground and excited states of both the diquark and quark-diquark bound systems are considered. Here the light quark is treated completely relativistically and quark-diquark potential is constructed. In the present study, we employed the theoretical approach called Regge Phenomenology with the assumption of linear Regge trajectories. Several relations between Regge intercepts, slope ratios, and baryon masses are extracted and the ground state masses of $\Xi_{cc}$, $\Xi_{bc}$, $\Omega_{cc}$, and $\Omega_{bc}$ baryons are evaluated using these equations. After that, we calculate the Regge slopes for $\frac{1}{2}^{+}$ and $\frac{3}{2}^{+}$ trajectories to obtain the excited state masses for natural and unnatural parity states in the ($J,M^{2}$) plane. Further we extend our work and extract the Regge parameters in the ($n,M^{2}$) and obtained the radial and orbital excited state masses of these doubly heavy baryons. 
	
	The remainder of this paper is organised as follows. After a brief introduction, in Sec II, the complete Regge theory is described and the mass relations are extracted. Using these relations, the ground and excited masses are estimated in the $(J,M^{2})$ plane. After that, we extend our work and determine the Regge parameters to obtain the mass spectra of these baryons in the ($n,M^{2}$) plane. Sec III provides the complete discussion of our results obtained. Finally, we concluded our work in section IV.

	\section{Theoretical Framework}
	
	The linear Regge trajectory is one of the most effective and extensively used phenomenological approaches to studying hadron spectroscopy. To understand the Regge trajectory, various theories were developed. In 1978, Nambu proposed one of the most straightforward explanation for linear Regge trajectories \cite{Nambu1974,Nambu1979}. He assumed that the uniform interaction of quark and antiquark pair creates a strong flux tube, with light quarks rotating with the speed of light at  radius $R$ at the tube's end. The mass originating in this flux tube is estimated as \cite{Nambu1974,Nambu1979,Oudichhya},
	\begin{equation}
		\label{eq1}
		M = 2\int_{0}^{R}\frac{\sigma}{\sqrt{1-\nu^{2}(r)}}dr = \pi\sigma R ,
	\end{equation}
	where $\sigma$ denotes the string tension (the mass density per unit length). Similarly, the flux tube's angular momentum is calculated as,
	\begin{equation}
		\label{eq2}
		J = 2\int_{0}^{R}\frac{\sigma r \nu(r)}{\sqrt{1-\nu^{2}(r)}}dr = \frac{\pi\sigma R^{2}}{2}+c^{'} ,
	\end{equation}
	one can also write,
	\begin{equation}
		\label{eq3}
		J = \frac{M^{2}}{2\pi\sigma} + c^{''} ,
	\end{equation}
	where $c^{'}$ and $c^{''}$ are the constants of integration. Hence we can say that, $J$ and $M^{2}$ are linearly related to each other. The plots of Regge trajectories of hadrons in the $(J,M^{2})$ plane are usually called Chew-Frautschi plots \cite{Chew1961}. They used the theory to investigate the strong quark gluon interaction and observed that the experimentally missing excited states of mesons and baryons lie on the linear trajectories in the  $(J,M^{2})$ plane. A set of internal quantum numbers characterises a particular pole's trajectory, and the hadrons lying on the same Regge line have the same internal quantum numbers. The general relation for the linear Regge trajectories can be expressed as \cite{Nambu1974,Nambu1979,Oudichhya,Wei2008},
	\begin{eqnarray}
		\label{eq4}
		J = \alpha(M) = a(0)+\alpha^{'} M^{2} ,
	\end{eqnarray}
	where $a(0)$ and $\alpha^{'}$ are, respectively, the slope and intercept
	of the Regge trajectory. In reality, the linear property in Regge theory is just the leading order phenomenon. The Regge trajectories have been demonstrated to be nonlinear in several studies \cite{Brisudova2000,Zhang A2005,Wei2013,Add1}. Generally the nonlinear behaviour of Regge trajectories are observed in quarkonia \cite{Kher2022,Raghav1,Raghav2,Raghav3}. Although string-like or semi relativistic potential models may show a linear relationship between hadron mass squared and quantum numbers.
	
	 Now for different flavors of baryon multiplets, the Regge parameters ($\alpha^{'}$ and $a(0)$) can be related by the following relations \cite{Wei2008,Juhi,Add2,Add3,A.B.1982},
	 \\
	 the additivity of intercepts
	\begin{eqnarray}
		\label{eq5}
		a_{iiq}(0) + a_{jjq}(0) = 2a_{ijq}(0) ,	
	\end{eqnarray}
	the additivity of inverse slopes
	\begin{eqnarray}
		\label{eq6}
		\frac{1}{{\alpha^{'}}_{iiq}} + \frac{1}{{\alpha^{'}}_{jjq}} = \frac{2}{{\alpha^{'}}_{ijq}} ,	
	\end{eqnarray}\\
	where $i, j$, and $q$ represents the quark flavors. Using Eqs. (\ref{eq4}) and (\ref{eq5}) we obtain,
	\begin{eqnarray}
		\label{eq7}
		\alpha^{'}_{iiq}M^{2}_{iiq}+\alpha^{'}_{jjq}M^{2}_{jjq}=2\alpha^{'}_{ijq}M^{2}_{ijq} .
	\end{eqnarray}
	\\	
	After combining the Eqs. (\ref{eq6}) and (\ref{eq7}), we derive two pairs of solutions which are expressed as,
	
	\begin{eqnarray}
		\label{eq8}
		\fl \frac{\alpha^{'}_{jjq}}{\alpha^{'}_{iiq}}=\frac{1}{2M^{2}_{jjq}}\times[(4M^{2}_{ijq}-M^{2}_{iiq}-M^{2}_{jjq})
		\pm\sqrt{{{(4M^{2}_{ijq}-M^{2}_{iiq}-M^{2}_{jjq}})^2}-4M^{2}_{iiq}M^{2}_{jjq}}],
	\end{eqnarray}
	and, 
	\begin{eqnarray}
		\label{eq9}	
		\fl	\frac{\alpha^{'}_{ijq}}{\alpha^{'}_{iiq}}=\frac{1}{4M^{2}_{ijq}}\times[(4M^{2}_{ijq}+M^{2}_{iiq}-M^{2}_{jjq})
		\pm\sqrt{{{(4M^{2}_{ijq}-M^{2}_{iiq}-M^{2}_{jjq}})^2}-4M^{2}_{iiq}M^{2}_{jjq}}].
	\end{eqnarray}
	\\
	These are the important relations that we have derived between slope ratios and baryon masses.
	Now the above obtained Eq. (\ref{eq8}) can also be expressed as,
	\begin{eqnarray}
		\label{eq10}
		\frac{\alpha^{'}_{jjq}}{\alpha^{'}_{iiq}}=	\frac{\alpha^{'}_{kkq}}{\alpha^{'}_{iiq}}\times	\frac{\alpha^{'}_{jjq}}{\alpha^{'}_{kkq}} ,
	\end{eqnarray}	
	here $k$ can be any quark flavor. As a result,
	
	\begin{eqnarray}
		\label{eq11}
		\nonumber
		\fl	&\frac{[(4M^{2}_{ijq}-M^{2}_{iiq}-M^{2}_{jjq})+\sqrt{{{(4M^{2}_{ijq}-M^{2}_{iiq}-M^{2}_{jjq}})^2}-4M^{2}_{iiq}M^{2}_{jjq}}]}{2M^{2}_{jjq}} \\ 
		\fl	&=\frac{[(4M^{2}_{ikq}-M^{2}_{iiq}-M^{2}_{kkq})+\sqrt{{{(4M^{2}_{ikq}-M^{2}_{iiq}-M^{2}_{kkq}})^2}-4M^{2}_{iiq}M^{2}_{kkq}}]/2M^{2}_{kkq}}{[(4M^{2}_{jkq}-M^{2}_{jjq}-M^{2}_{kkq})+\sqrt{{{(4M^{2}_{jkq}-M^{2}_{jjq}-M^{2}_{kkq}})^2}-4M^{2}_{jjq}M^{2}_{kkq}}]/2M^{2}_{kkq}} . 
	\end{eqnarray}
	\\
	The above equation obtained in terms of baryon masses is the general relation, which can be used to determine the mass of any baryon state if all other masses are known. According to a number of theoretical investigations, the slope of Regge trajectories decreases as quark masses increases \cite{Brisudova2000,Zhang A2005,Add1,Add2,A.B.1982,J.L.1986}. For $m_{j}>m_{i}$, we can write $\alpha^{'}_{jjq}/\alpha^{'}_{iiq}<1$. As a result of Eq. (\ref{eq8}), we can say that,
	
	\begin{eqnarray}
		\label{eq12} \nonumber
		\frac{1}{2M^{2}_{jjq}}\times[(4M^{2}_{ijq}-M^{2}_{iiq}-M^{2}_{jjq}) \\ 
		+\sqrt{{{(4M^{2}_{ijq}-M^{2}_{iiq}-M^{2}_{jjq}})^2}-4M^{2}_{iiq}M^{2}_{jjq}}] < 1 ,
	\end{eqnarray}
	the above relation gives the inequality equation which is expressed as,
	\begin{equation}
		\label{eq13}
		2M^{2}_{ijq}<M^{2}_{iiq}+M^{2}_{jjq} .
	\end{equation}
	We now introduce a parameter called $\delta$, which is used to estimate the deviation of a relation (\ref{eq13}) by replacing the sign of inequality with the equal sign. For baryons, it is represented by $\delta^{b}_{ij,q}$ and given by,
	\begin{equation}
		\label{eq14}
		\delta^{b}_{ij,q} = M^{2}_{iiq}+M^{2}_{jjq}-2M^{2}_{ijq} ,
	\end{equation}
	here also $i, j$, and $q$ represent the arbitrary light or heavy quarks.
	Now from Eqs. (\ref{eq5}) and (\ref{eq6}) we can write,
	\begin{equation}
		\label{eq15}
		a_{iiq}(0)-a_{ijq}(0)=a_{ijq}(0)-a_{jjq}(0),
	\end{equation}
	
	\begin{equation}
		\label{eq16}
		\frac{1}{\alpha_{iiq}^{'}}-\frac{1}{\alpha_{ijq}^{'}}=	\frac{1}{\alpha_{ijq}^{'}}-\frac{1}{\alpha_{jjq}^{'}} ,
	\end{equation}
	we introduce two parameters based on above equations \cite{Wei2008},
	\begin{equation}
		\label{eq17}
		\lambda_{x}=a_{nnn}(0)-a_{nnx}(0)  ,  \gamma_{x}=\frac{1}{\alpha^{'}_{nnx}}-\frac{1}{\alpha^{'}_{nnn}} ;
	\end{equation}
	where $n$ represents light nonstrange quark ($u$ or $d$) and $x$ denotes $i, j$, or $q$. From Eqs. (\ref{eq15})-(\ref{eq17} )we have,
	
	\begin{eqnarray}
		\label{eq18}
		a_{ijq}(0) = a_{nnn}(0)-\lambda_{i}-\lambda_{j}-\lambda_{q} , \frac{1}{\alpha_{ijq}^{'}} = \frac{1}{\alpha_{nnn}^{'}}+\gamma_{i}+\gamma_{j}+\gamma_{q}.
	\end{eqnarray}
	Since in baryon multiplets for $nnn$ and $ijq$ states, we can write from Eq. (\ref{eq4}),
	\begin{eqnarray}
		\label{eq19} \nonumber
		J &= a_{nnn}(0) + \alpha_{nnn}^{'}M^{2}_{nnn} ,\\ 
		J &= a_{ijq}(0) + \alpha^{'}_{ijq}M^{2}_{ijq} ,
	\end{eqnarray}
	solving Eqs. (\ref{eq18}) and (\ref{eq19}) we have,
	\begin{equation}
		\label{eq20}
		M^{2}_{ijq} = (\alpha^{'}_{nnn}M^{2}_{nnn} + \lambda_{i} + \lambda_{j} + \lambda_{q})\left(\frac{1}{\alpha^{'}_{nnn}}+\gamma_{i}+\gamma_{j}+\gamma_{q}\right).
	\end{equation}
	Now combining relations (\ref{eq14}) and (\ref{eq20}) we can prove that \cite{Oudichhya,Wei2008},
	\begin{eqnarray}
		\label{eq21}
		\nonumber
		\delta^{b}_{ij,q} &=& M^{2}_{iiq}+M^{2}_{jjq}-2M^{2}_{ijq}\\ 
		&=& 2(\lambda_{i}-\lambda_{j})(\gamma_{i}-\gamma_{j}) ,
	\end{eqnarray}
	which says that $\delta^{b}_{ij,q}$ is independent of quark flavor $q$.
	
	\subsection{Ground state masses of $\Xi_{cc}$, $\Xi_{bc}$, $\Omega_{cc}$, and $\Omega_{bc}$ baryons.}
	
	In this section, we compute the ground state ($J^{P}=\frac{1}{2}^{+}$ and $\frac{3}{2}^{+}$) masses of $\Xi_{cc}$, $\Xi_{bc}$, $\Omega_{cc}$, and $\Omega_{bc}$ baryons using the relations we have extracted above. For doubly charmed baryons, in the case of $\Xi_{cc}^{++}$ which is composed of ($ucc$), we insert $i=d$, $j=c$, $q=u$, and $k=s$ in Eq. (\ref{eq11}) and obtain the mass expression for $\Xi_{cc}^{++}$  as a function of well established light baryon masses and singly charmed baryon masses which is expressed as, 
	\\
	
	\begin{eqnarray}
		\label{eq22}
		\nonumber
		\fl &\frac{[(4M^{2}_{\Lambda_{c}^{+}}-M^{2}_{n}-M^{2}_{\Xi_{cc}^{++}})+\sqrt{(4M^{2}_{\Lambda_{c}^{+}}-M^{2}_{n}-M^{2}_{\Xi_{cc}^{++}})^{2}-4M^{2}_{n}M^{2}_{\Xi_{cc}^{++}}}]}{2M^{2}_{\Xi_{cc}^{++}}}\\ 
		&=\frac{[(4M^{2}_{\Sigma^{0}}-M^{2}_{n}-M^{2}_{\Xi^{0}})+\sqrt{(4M^{2}_{\Sigma^{0}}-M^{2}_{n}-M^{2}_{\Xi^{0}})^{2}-4M^{2}_{n}M^{2}_{\Xi^{0}}}]}{[(4M^{2}_{\Xi_{c}^{+}}-M^{2}_{\Xi_{cc}^{++}}-M^{2}_{\Xi^{0}})+\sqrt{(4M^{2}_{\Xi_{c}^{+}}-M^{2}_{\Xi_{cc}^{++}}-M^{2}_{\Xi^{0}})^{2}-4M^{2}_{\Xi_{cc}^{++}}M^{2}_{\Xi^{0}}}]} \ \ \ \ \
	\end{eqnarray}
	
	Similarly we can get the mass expression for  $\Omega_{cc}^{*+}$ baryon which is comprises of one strange quark ($s$) and two charm quarks ($c$), hence by putting $i=d$, $j=s$, $q=s$, and $k=c$ in Eq. (\ref{eq11}) we have,
	\begin{eqnarray}
		\label{eq23}
		\nonumber
		\fl &\frac{[(4M^{2}_{\Xi^{*-}}-M^{2}_{\Sigma^{*-}}-M^{2}_{\Omega^{-}})+\sqrt{(4M^{2}_{\Xi^{*-}}-M^{2}_{\Sigma^{*-}}-M^{2}_{\Omega^{-}})^{2}-4M^{2}_{\Sigma^{*-}}M^{2}_{\Omega^{-}}}]}{2M^{2}_{\Omega^{-}}}\\ 
		&=\frac{[(4M^{2}_{\Xi_{c}^{*0}}-M^{2}_{\Sigma^{*-}}-M^{2}_{\Omega_{cc}^{*+}})+\sqrt{(4M^{2}_{\Xi_{c}^{*0}}-M^{2}_{\Sigma^{*-}}-M^{2}_{\Omega_{cc}^{*+}})^{2}-4M^{2}_{\Sigma^{*-}}M^{2}_{\Omega_{cc}^{*+}}}]}{[(4M^{2}_{\Omega_{c}^{*0}}-M^{2}_{\Omega^{-}}-M^{2}_{\Omega_{cc}^{*+}})+\sqrt{(4M^{2}_{\Omega_{c}^{*0}}-M^{2}_{\Omega^{-}}-M^{2}_{\Omega_{cc}^{*+}})^{2}-4M^{2}_{\Omega_{cc}^{*+}}M^{2}_{\Omega^{-}}}]} \ \ \ \ \ 
	\end{eqnarray}
	\\
	The above relation (\ref{eq23}) we obtained for $\Omega_{cc}^{*+}$ is in terms of well known light baryon masses and singly charmed baryon masses.
	Hence, after inserting the masses of $\Lambda_{c}^{+}$, neutron ($n$), $\Sigma^{0}$, $\Xi^{0}$, $\Xi_{c}^{+}$, $\Xi^{*-}$, $\Sigma^{*-}$, $\Omega^{-}$, $\Xi_{c}^{*0}$ from PDG \cite{PDG} and mass of $\Omega_{c}^{*0}$ from \cite{Oudichhya} (calculated in previous work) in Eqs. (\ref{eq22}) and (\ref{eq23}), we get $M_{\Xi_{cc}^{++}}$ = 3.579 GeV for $J^{P}=\frac{1}{2}^{+}$  and $M_{\Omega_{cc}^{*+}}$ = 3.847 GeV for $J^{P}=\frac{3}{2}^{+}$. Since, only the charge mixed states of $\Delta(1232)$ were assuredly measured as mentioned in PDG \cite{PDG}, therefore we avoided using the masses of $\Delta^{++}$, $\Delta^{+}$, $\Delta^{0}$, and $\Delta^{-}$ baryons in this computation. Now, $\delta^{b}_{ij,q}$ is independent of the quark flavor $q$, as already stated in the above relation (\ref{eq21}). As a result, using Eq. (\ref{eq14}) we can have the following relations:
	\\
	(I) $i=u$, $j=s$, $q=n$, $c$
	\begin{eqnarray}
		\label{eq24}
		\nonumber	
		\delta_{us}^{(3/2)^{+}}&=&M^{2}_{\Delta}+M^{2}_{\Xi^{*}}-2M^{2}_{\Sigma^{*}}\\
		&=&M^{2}_{\Sigma_{c}^{*}}+M^{2}_{\Omega_{c}^{*}}-2M^{2}_{\Xi_{c}^{*}};
	\end{eqnarray}
	(II) $i=u$, $j=c$, $q=n$ ,$s$
	\begin{eqnarray}
		\label{eq25}
		\nonumber
		\delta_{uc}^{(3/2)^{+}}&=&M^{2}_{\Delta}+M^{2}_{\Xi^{*}_{cc}}-2M^{2}_{\Sigma^{*}_{c}}\\
		&=&M^{2}_{\Sigma^{*}}+M^{2}_{\Omega_{cc}^{*}}-2M^{2}_{\Xi_{c}^{*}};
	\end{eqnarray}
	(III) $i=c$, $j=b$, $q=n$ ,$s$
	\begin{eqnarray}
		\label{eq26}
		\nonumber
		\delta_{cb}^{(3/2)^{+}}&=&M^{2}_{\Xi_{cc}^{*}}+M^{2}_{\Xi^{*}_{bb}}-2M^{2}_{\Xi^{*}_{bc}}\\
		&=&M^{2}_{\Omega_{cc}^{*}}+M^{2}_{\Omega_{bb}^{*}}-2M^{2}_{\Omega_{bc}^{*}};
	\end{eqnarray}
	solving Eqs. (\ref{eq24}) and (\ref{eq25}) we have,
	\begin{equation}
		\label{eq27}
		(M^{2}_{\Omega_{cc}^{*}}-M^{2}_{\Xi^{*}_{cc}})+(M^{2}_{\Xi^{*}}-M^{2}_{\Sigma^{*}}) = (M^{2}_{\Omega_{c}^{*}}-M^{2}_{\Sigma_{c}^{*}}),
	\end{equation}
	similarly, its corresponding relation for $\frac{1}{2}^{+}$ multiplet is expressed as,
	\begin{equation}
		\label{eq28}
		(M^{2}_{\Omega_{cc}}-M^{2}_{\Xi_{cc}})	+ (M^{2}_{\Xi}-M^{2}_{\Sigma}) = (M^{2}_{\Omega_{c}}-M^{2}_{\Sigma_{c}}).
	\end{equation}
	Using Eq. (\ref{eq27}), the mass expressions for $\Xi_{cc}^{*++}$ and $\Xi_{cc}^{*+}$ baryons can be obtained and again after inserting the masses of $\Xi^{*}$, $\Sigma^{*}$, and $\Sigma_{c}^{*}$ from PDG \cite{PDG} and $\Omega_{c}^{*}$ and $\Omega_{cc}^{*}$ (calculated in our work), we get  $M_{\Xi_{cc}^{*++}}$ = 3.729 GeV and $M_{\Xi_{cc}^{*+}}$ = 3.726 GeV for $J^{P} = \frac{3}{2}^{+}$ state. Similarly, with the help of Eq. (\ref{eq28}), we can get $M_{\Omega_{cc}^{+}}$ = 3.719 GeV for $J^{P} = \frac{1}{2}^{+}$ state. Now for the bottom-charm system $\Xi_{bc}^{+}$ and $\Xi_{bc}^{0}$, we again use Eq. (\ref{eq11}) to evaluate the ground state masses. For $\Xi_{bc}^{0}$ baryon ($dcb$) containing two different heavy quarks and one light quark, we put $i=c$, $j=b$, $q=d$, and $k=c$ into relation (\ref{eq11}) and after simplifying  we get, 
	
	\begin{eqnarray}
		\label{eq29}
		\fl	[(M_{\Xi^{+}_{cc}} + M_{\Xi_{bb}^{-}})^{2} - 4M^{2}_{\Xi_{bc}^{0}}] = \sqrt{(4M^{2}_{\Xi_{bc}^{0}}-M^{2}_{\Xi^{+}_{cc}}-M^{2}_{\Xi_{bb}^{-}})^{2}-4M^{2}_{\Xi^{+}_{cc}}M^{2}_{\Xi_{bb}^{-}}}
	\end{eqnarray}
	\\
	In the quadratic mass relation we have derived above for $\Xi_{bc}^{0}$ baryon, after
		inserting the masses of $\Xi_{cc}^{+}$ (calculated above) and $\Xi_{bb}^{-}$ (calculated in previous work \cite{Juhi}) we get the ground state mass of $\Xi_{bc}^{0}$ baryon as 6.906 GeV for $J^{P} = \frac{1}{2}^{+}$ state. Similarly we obtained $M_{\Xi_{bc}^{+}}$ = 6.902 GeV, $M_{\Xi_{bc}^{*0}}$ = 7.029 GeV, and $M_{\Xi_{bc}^{*+}}$ = 7.030 GeV. In the same manner for $\Omega_{bc}^{0}$ baryon composed of two different heavy flavors and one light strange quark, with the aid of relation (\ref{eq26}) we can derive the mass expression for $\Omega_{bc}^{0}$ and after putting the values of masses of $\Xi_{cc}$, $\Omega_{cc}$, $\Xi_{bc}$ (calculated above) and $\Xi_{bb}$ (calculated in previous work \cite{Juhi}),  we evaluated the ground state masses of $\Omega_{bc}^{0}$ baryon as 7.035 GeV and 7.149 GeV for $J^{P} = \frac{1}{2}^{+}$ and $\frac{3}{2}^{+}$  states respectively.

	\begin{table}
		\begin{center}
			
			\caption{\label{table1}Masses of excited states of the $\Xi_{cc}$ baryon in the $(J,M^{2})$ plane for natural and unnatural parity states (in GeV).}
			\footnotesize
			\begin{tabular}{@{}lllllllllll}
				\br
				\textit{$N^{2S+1}L_{J}$}&\multicolumn{2}{c}{Present}& & &Others	\\
				\hline	\noalign{\smallskip}
				&$\Xi_{cc}^{++}$ &$\Xi_{cc}^{+}$ & PDG \cite{PDG}  & \cite{Yoshida2015} & \cite{Roberts2008} & \cite{Ebert2002}  & \cite{Wei2015} & \cite{S. Gershteincas} &\cite{Rosner2014}& \cite{Alexandrou}\\	
				\mr
				
				$1^{2}S_{\frac{1}{2}}$ &  3.579 &3.581 &3.621& 3.685 &3.676& 3.620&3.520 &3.478 &3.627 &3.568\\
				$1^{2}P_{\frac{3}{2}}$ &  3.924 &3.926 & &3.949 &3.921 &3.959 &3.786 &3.834\\
				$1^{2}D_{\frac{5}{2}}$ &  4.241 &4.243 & &4.115 &4.047 & &4.034 &4.050\\
				$1^{2}F_{\frac{7}{2}}$ &  4.536 &4.538 & & & & &4.267\\
				$1^{2}G_{\frac{9}{2}}$ &  4.813 &4.815 & &\\
				$1^{2}H_{\frac{11}{2}}$&  5.075 &5.077 & &\\
				\noalign{\smallskip}
				$1^{4}S_{\frac{3}{2}}$ &  3.729 &3.726 & &3.754 &3.753 &3.727 &3.695 &3.610 &3.690 &3.652\\
				$1^{4}P_{\frac{5}{2}}$ &  4.076 &4.074 & &4.163 &4.092 &4.155 & 3.949 &4.047\\ 
				$1^{4}D_{\frac{7}{2}}$ &  4.396 &4.394 & & & 4.097 & &4.187 &4.089\\
				$1^{4}F_{\frac{9}{2}}$ &  4.694 &4.692 & & & & & 4.413\\
				$1^{4}G_{\frac{11}{2}}$&  4.974 &4.973 & &\\
				$1^{4}H_{\frac{13}{2}}$&  5.240 &5.239 & &\\
				\br
			\end{tabular}\\
		\end{center}
	\end{table}
	\normalsize
	
	\begin{table}
		\begin{center}
			
			\caption{\label{table2}Masses of excited states of the $\Omega_{cc}$ baryon in the $(J,M^{2})$ plane for natural and unnatural parity states (in GeV).}
			\footnotesize
			\begin{tabular}{@{}lllllllllll}
				\br
				\textit{$N^{2S+1}L_{J}$}&Present& \cite{ZalakOmega_cc} & \cite{Ebert2002} & \cite{Wei2015} & \cite{Roberts2008} & \cite{S. Migura2006}	\\
				
				\mr
				
				$1^{2}S_{\frac{1}{2}}$ & 3.719 &3.650 &3.778 &3.650 &3.815 &3.732 \\
				$1^{2}P_{\frac{3}{2}}$ & 3.947 &3.972 &4.102 &3.910 &4.052 &3.986 \\
				$1^{2}D_{\frac{5}{2}}$ & 4.162 &4.141 & &4.153 &4.202 \\
				$1^{2}F_{\frac{7}{2}}$ & 4.367 &4.387 & &4.383 \\
				$1^{2}G_{\frac{9}{2}}$ & 4.562 \\
				$1^{2}H_{\frac{11}{2}}$& 4.749 \\
				\noalign{\smallskip}
				$1^{4}S_{\frac{3}{2}}$ & 3.847 &3.810 &3.872 &3.809 &3.876 &3.765 \\
				$1^{4}P_{\frac{5}{2}}$ & 4.153 &3.958 &4.303 &4.058 &4.152\\ 
				$1^{4}D_{\frac{7}{2}}$ & 4.438 &4.122 & &4.294 &4.230\\
				$1^{4}F_{\frac{9}{2}}$ & 4.706 & 4.274 &4.516\\
				$1^{4}G_{\frac{11}{2}}$& 4.960 \\
				$1^{4}H_{\frac{13}{2}}$& 5.201 \\
				\br
			\end{tabular}\\
		\end{center}
	\end{table}
	\normalsize
	
	\begin{table}
		\begin{center}
			
			\caption{\label{table3}Masses of excited states of the $\Xi_{bc}$ baryon in the $(J,M^{2})$ plane for natural and unnatural parity states (in GeV).}
			\footnotesize
			\begin{tabular}{@{}lllllllllll}
				\br
				\textit{$N^{2S+1}L_{J}$}&\multicolumn{2}{c}{Present}& & Others	\\
				\hline	\noalign{\smallskip}
				&$\Xi_{bc}^{+}$ &$\Xi_{bc}^{0}$ & \cite{B. Eakins2012}  & \cite{Roberts2008} & \cite{Giannuzzi2009}   & \cite{Z.S. Brown2014} &\cite{Ebert2002} & \cite{Rosner2014} &\cite{Mathur N2018}\\	
				\mr
				
				$1^{2}S_{\frac{1}{2}}$ & 6.902 &6.906 &7.014 &7.011 &6.904 &6.943 &6.933 &6.914&6.945\\
				$1^{2}P_{\frac{3}{2}}$ & 7.254 &7.258 &7.394\\
				$1^{2}D_{\frac{5}{2}}$ & 7.590 &7.594 \\
				$1^{2}F_{\frac{7}{2}}$ & 7.912 &7.916 \\
				$1^{2}G_{\frac{9}{2}}$ & 8.221 &8.225 \\
				$1^{2}H_{\frac{11}{2}}$& 8.519 &8.523 \\
				\noalign{\smallskip}
				$1^{4}S_{\frac{3}{2}}$ & 7.030 &7.029 & 7.064 &7.074 &6.936 &6.985 &6.980 &6.969&6.989\\
				$1^{4}P_{\frac{5}{2}}$ & 7.378 &7.377  \\ 
				$1^{4}D_{\frac{7}{2}}$ & 7.710 &7.709 &  \\
				$1^{4}F_{\frac{9}{2}}$ & 8.029 &8.028 &\\
				$1^{4}G_{\frac{11}{2}}$& 8.335 &8.334  \\
				$1^{4}H_{\frac{13}{2}}$& 8.631 &8.630  \\	
				\br
			\end{tabular}\\

		\end{center}
	\end{table}
	\normalsize
	
	\begin{table}
		\begin{center}
			
			\caption{\label{table4}Masses of excited states of the $\Omega_{bc}$ baryon in the $(J,M^{2})$ plane for natural and unnatural parity states (in GeV).}
			\footnotesize
			\begin{tabular}{@{}lllllllllll}
				\br
				\textit{$N^{2S+1}L_{J}$}&Present& \cite{ZalakOmega_cc} &\cite{Ebert2002} &\cite{Giannuzzi2009} &\cite{Roberts2008} & \cite{Roncaglia1995} & \cite{Mathur N2018}	\\
				
				\mr
				
				$1^{2}S_{\frac{1}{2}}$ & 7.035 & 7.136 &7.088 &6.994 &7.136 &7.045 &6.994\\
				$1^{2}P_{\frac{3}{2}}$ & 7.261 &7.373\\
				$1^{2}D_{\frac{5}{2}}$ & 7.480 &7.547\\
				$1^{2}F_{\frac{7}{2}}$ & 7.693 &7.705\\
				$1^{2}G_{\frac{9}{2}}$ & 7.900 \\
				$1^{2}H_{\frac{11}{2}}$& 8.102 \\
				\noalign{\smallskip}
				$1^{4}S_{\frac{3}{2}}$ & 7.149 &7.187 &7.130 &7.017 &7.187 &7.120 &7.056 \\
				$1^{4}P_{\frac{5}{2}}$ & 7.452 &7.363\\ 
				$1^{4}D_{\frac{7}{2}}$ & 7.744 &7.534\\
				$1^{4}F_{\frac{9}{2}}$ & 8.025 &7.690\\
				$1^{4}G_{\frac{11}{2}}$& 8.296 \\
				$1^{4}H_{\frac{13}{2}}$& 8.559 \\
				\br
			\end{tabular}\\
		\end{center}
	\end{table}
	\normalsize

	\subsection{Excited state masses in the ($J,M^{2}$) plane.}
	
	After calculating the ground-state masses, in this section the excited state masses of doubly heavy $\Xi$ and $\Omega$ baryons lying on $\frac{1}{2}^{+}$ and $\frac{3}{2}^{+}$ trajectories are obtained and for that we firstly extract the Regge slopes $\alpha^{'}$. For instance, to calculate $\alpha^{'}_{ncc}$ we put $i=s$, $j=c$, and $q=n (u$ or $d)$ in Eq. (\ref{eq8}) we have,
	
	\begin{eqnarray}
		\label{eq30}
		\fl	\frac{\alpha^{'}_{ncc}}{\alpha^{'}_{nss}}=\frac{1}{2M^{2}_{\Xi_{cc}}}\times[(4M^{2}_{\Xi_{c}}-M^{2}_{\Xi}-M^{2}_{\Xi_{cc}})
		+\sqrt{{{(4M^{2}_{\Xi_{c}}-M^{2}_{\Xi}-M^{2}_{\Xi_{cc}})^2}-4M^{2}_{\Xi}M^{2}_{\Xi_{cc}}}}],
	\end{eqnarray}
	\\
	After inserting the masses of $\Xi$ and $\Xi_{c}$ from PDG \cite{PDG} and $\Xi_{cc}$ (calculated above) for $J^{P}=\frac{1}{2}^{+}$ in the above equation, we get $\alpha^{'}_{ncc}/\alpha^{'}_{nss}$. The value of $\alpha^{'}_{nss}$ is calculated as 0.7420 GeV$^{-2}$  in the similar manner by putting the values  $i=n$, $j=s$ and $q=n$ in Eq. (\ref{eq8}).
		Also from slope equation (\ref{eq4}) we can have,
		\begin{equation}
			\nonumber
			\alpha^{'} = \frac{(J+2)-J}{M^{2}_{J+2}-M^{2}_{J}} ,
	\end{equation} 
	Hence, using this equation we have $\alpha^{'}_{ncc}$ = 0.3866 GeV$^{-2}$ for $\frac{1}{2}^{+}$ trajectory. Similarly with the help of Eqs. (\ref{eq8}) and (\ref{eq9}), we can determine the values of $\alpha^{'}_{scc}$, $\alpha^{'}_{ncb}$, and $\alpha^{'}_{scb}$ for both $\frac{1}{2}^{+}$ and $\frac{3}{2}^{+}$ trajectories. 	Using Eq. (\ref{eq4}) one can write,
	\begin{eqnarray}
		\label{eq31}
		M_{J+1} = \sqrt{M_{J}^{2}+\frac{1}{\alpha^{'}}} .
	\end{eqnarray}
	With the help of values of $\alpha^{'}$ extracted, from Eq. (\ref{eq31}) we can obtain the excited state masses of $\Xi_{cc}$, $\Xi_{bc}$, $\Omega_{cc}$, and $\Omega_{bc}$ baryons with  $J^{P}= \frac{3}{2}^{-}, \frac{5}{2}^{+}, \frac{7}{2}^{-}$.... and $J^{P}= \frac{5}{2}^{-}, \frac{7}{2}^{+}, \frac{9}{2}^{-}$.... for natural and unnatural parities respectively, in the $(J,M^{2})$ plane. Our calculated results are shown in Tables \ref{table1}-\ref{table4} with other theoretical predictions. Here the state of the particles is represented by the spectroscopic notations \textit{$N^{2S+1}L_{J}$}, where $N$, $L$, and $S$ denote the radial excited quantum number, orbital quantum number, and intrinsic spin, respectively.
	
	\begin{table}
		\begin{center}
			
			\caption{\label{table5}Masses of excited states of the $\Xi_{cc}$ baryon in the $(n,M^{2})$ plane. The masses in the boldface are taken as input from Ref. \cite{Zalakcas_cc}  (in GeV).}
			\footnotesize
			\begin{tabular}{@{}lllllllllll}
				\br
				&	\textit{$N^{2S+1}L_{J}$}&\multicolumn{2}{c}{Present}& & Others	\\
				\hline\noalign{\smallskip}
				& &$\Xi_{cc}^{++}$ & $\Xi_{cc}^{+}$ & \cite{Yoshida2015} & \cite{Roberts2008}& \cite{Ebert2002} &\cite{B. Eakins2012} &\cite{Wei2015} \\
				\mr
				(S=1/2)& $1^{2}S_{1/2}$ &3.579 & 3.581 &3.685 &3.676 &3.620 &3.678 &3.520\\
				& $2^{2}S_{1/2}$ &\textbf{3.920} &\textbf{3.925} &4.079 &4.029 &3.910 &4.030\\
				& $3^{2}S_{1/2}$ &4.234 &4.241 &4.206 & &4.154\\
				& $4^{2}S_{1/2}$ &4.526 &4.535\\
				& $5^{2}S_{1/2}$ &4.800 &4.812\\
				& $6^{2}S_{1/2}$ &5.059 &5.073\\
				
				(S=3/2)& $1^{4}S_{3/2}$ &3.729 &3.726 &3.754 &3.753 &3.727 &3.752 &3.695\\
				& $2^{4}S_{3/2}$ &\textbf{3.983} &\textbf{3.988} &4.114 &4.042 &4.027 &4.078\\
				& $3^{4}S_{3/2}$ &4.223 &4.234&4.131\\
				& $4^{4}S_{3/2}$ &4.448 &4.466\\
				& $5^{4}S_{3/2}$ &4.663 &4.687\\
				& $6^{4}S_{3/2}$ &4.868 &4.898\\   
				
				\noalign{\smallskip}\hline\noalign{\smallskip}    
				
				(S=1/2)& $1^{2}P_{3/2}$ &3.924 &3.926 &3.949 &3.921 &3.959 &4.079 &3.786\\
				& $2^{2}P_{3/2}$ &\textbf{4.140}&\textbf{4.144} &4.137 &4.078\\
				& $3^{2}P_{3/2}$ &4.345 &4.351 &4.159\\
				& $4^{2}P_{3/2}$ &4.541 &4.549\\
				& $5^{2}P_{3/2}$ &4.729 &4.738\\
				
				(S=3/2) & $1^{4}P_{5/2}$ &4.076 &4.074 &4.163 &4.092 &4.155 &4.047 &3.949\\
				& $2^{4}P_{5/2}$ &\textbf{4.181} &\textbf{4.183} &4.488\\
				& $3^{4}P_{5/2}$ &4.283 &4.289 &4.534\\
				& $4^{4}P_{5/2}$ &4.383 &4.393\\
				& $5^{4}P_{5/2}$ &4.481 &4.494\\
				
				\noalign{\smallskip}\hline\noalign{\smallskip}    
				
				(S=1/2) & $1^{2}D_{5/2}$ &4.241 &4.243 &4.115 &4.047 & &4.050 &4.034\\
				& $2^{2}D_{5/2}$ &\textbf{4.297} &\textbf{4.299} &4.164 &4.091\\
				& $3^{2}D_{5/2}$ &4.352 &4.354 &4.348\\
				& $4^{2}D_{5/2}$ &4.407 &4.409\\
				& $5^{2}D_{5/2}$ &4.461 &4.463\\

				\br
			\end{tabular}\\
		\end{center}
	\end{table}
	\normalsize

	\begin{table}
		\begin{center}
			
			\caption{\label{table6}Masses of excited states of the $\Omega_{cc}$ baryon in the $(n,M^{2})$ plane. The masses in the boldface are taken as input from Ref. \cite{ZalakOmega_cc}  (in GeV).}
			\footnotesize
			\begin{tabular}{@{}lllllllllll}
				\br
				&	\textit{$N^{2S+1}L_{J}$} & Present & \cite{Ebert2002} & \cite{Yoshida2015} & \cite{Roberts2008} &\cite{Wei2015} &\\
				\mr
				(S=1/2)& $1^{2}S_{1/2}$ &3.719 &3.778 &3.832 &3.815 &3.650\\
				& $2^{2}S_{1/2}$ &\textbf{4.041} &4.075 &4.227 & 4.180\\
				& $3^{2}S_{1/2}$ &4.339 &4.321 &4.295\\
				& $4^{2}S_{1/2}$ &4.618 &\\
				& $5^{2}S_{1/2}$ &4.881\\
				& $6^{2}S_{1/2}$ &5.131\\
				
				(S=3/2)& $1^{4}S_{3/2}$ &3.847 &3.872 &3.838 &3.876 & 3.809\\
				& $2^{4}S_{3/2}$ &\textbf{4.096} &4.174 &4.263 &4.188\\
				& $3^{4}S_{3/2}$ &4.331 & &4.265\\
				& $4^{4}S_{3/2}$ &4.553\\
				& $5^{4}S_{3/2}$ &4.765\\
				& $6^{4}S_{3/2}$ &4.969\\   
				
				\noalign{\smallskip}\hline\noalign{\smallskip}    
				
				(S=1/2)& $1^{2}P_{3/2}$ &3.947 &4.102 &4.086 &4.052 &3.910\\
				& $2^{2}P_{3/2}$ &\textbf{4.259} &4.345 &4.201 &4.140 &\\
				& $3^{2}P_{3/2}$ &4.550\\
				& $4^{2}P_{3/2}$ &4.823\\
				& $5^{2}P_{3/2}$ &5.081\\
				
				(S=3/2) & $1^{4}P_{5/2}$ &4.153 & &4.220 &4.152\\
				& $2^{4}P_{5/2}$ &\textbf{4.247}\\
				& $3^{4}P_{5/2}$ &4.339\\
				& $4^{4}P_{5/2}$ &4.429\\
				& $5^{4}P_{5/2}$ &4.517\\
				
				\noalign{\smallskip}\hline\noalign{\smallskip}    
				
				(S=1/2) & $1^{2}D_{5/2}$ &4.162 & &4.264 &4.202 &4.153\\
				& $2^{2}D_{5/2}$ &\textbf{4.407}\\
				& $3^{2}D_{5/2}$ &4.640\\
				& $4^{2}D_{5/2}$ &4.860\\
				& $5^{2}D_{5/2}$ &5.071\\

				\br
			\end{tabular}\\
			
		\end{center}
	\end{table}
	\normalsize
	
	\begin{table}
		\begin{center}
			
			\caption{\label{table7}Masses of excited states of the $\Xi_{bc}$ baryon in the $(n,M^{2})$ plane. The masses in the boldface are taken as input from Ref. \cite{Zalakcas_cc}  (in GeV).}
			\footnotesize
			\begin{tabular}{@{}lllllllllll}
				\br
				&	\textit{$N^{2S+1}L_{J}$}&\multicolumn{2}{c}{Present}& & Others	\\
				\hline\noalign{\smallskip}
				& &$\Xi_{bc}^{+}$ & $\Xi_{bc}^{0}$ & \cite{B. Eakins2012} & \cite{Giannuzzi2009}& \cite{Z.S. Brown2014} &\cite{Roberts2008} \\
				\mr
				(S=1/2)& $1^{2}S_{1/2}$ &6.902 &6.906 &7.014 &6.904 &6.943&7.011\\
				& $2^{2}S_{1/2}$ &\textbf{7.240} &\textbf{7.244} &7.321 &7.478\\
				& $3^{2}S_{1/2}$ &7.563 &7.567 & &7.904\\
				& $4^{2}S_{1/2}$ &7.873 &7.876\\
				& $5^{2}S_{1/2}$ &8.170 &8.174\\
				& $6^{2}S_{1/2}$ &8.458 &8.462\\
				
				(S=3/2)& $1^{4}S_{3/2}$ &7.030 &7.029&7.064 &6.936 &6.985 &7.074\\
				& $2^{4}S_{3/2}$ &\textbf{7.263} &\textbf{7.267} &7.353 &7.495\\
				& $3^{4}S_{3/2}$ &7.489 &7.497 & &7.917\\
				& $4^{4}S_{3/2}$ &7.708 &7.721\\
				& $5^{4}S_{3/2}$ &7.921 &7.938\\
				& $6^{4}S_{3/2}$ &8.128 &8.150\\   
				
				\noalign{\smallskip}\hline\noalign{\smallskip}    
				
				(S=1/2)& $1^{2}P_{3/2}$ &7.254 &7.258 &7.394\\
				& $2^{2}P_{3/2}$ &\textbf{7.412} &\textbf{7.415}\\
				& $3^{2}P_{3/2}$ &7.567 &7.569\\
				& $4^{2}P_{3/2}$ &7.718 &7.719\\
				& $5^{2}P_{3/2}$ &7.867 &7.867\\
				
				(S=3/2) & $1^{4}P_{5/2}$ &7.378 &7.377\\
				& $2^{4}P_{5/2}$ &\textbf{7.434} &\textbf{7.408}\\
				& $3^{4}P_{5/2}$ &7.489 &7.439\\
				& $4^{4}P_{5/2}$ &7.545 &7.470\\
				& $5^{4}P_{5/2}$ &7.599 &7.500\\
				
				\br
			\end{tabular}\\

		\end{center}
	\end{table}
	\normalsize
	
	\begin{table}
		\begin{center}
			
			\caption{\label{table8}Masses of excited states of the $\Omega_{bc}$ baryon in the $(n,M^{2})$ plane. The masses in the boldface are taken as input from Ref. \cite{ZalakOmega_cc}  (in GeV).}
			\footnotesize
			\begin{tabular}{@{}lllllllllll}
				\br
				&	\textit{$N^{2S+1}L_{J}$} & Present  &\cite{Ebert2002} &\cite{Giannuzzi2009} &\cite{Roberts2008} & \cite{Roncaglia1995} \\
				\mr
				(S=1/2)& $1^{2}S_{1/2}$ &7.035 &7.088 &6.994 &7.136 &7.045\\
				& $2^{2}S_{1/2}$ &\textbf{7.480} & &7.559\\
				& $3^{2}S_{1/2}$ &7.900 & &7.976\\
				& $4^{2}S_{1/2}$ &8.299 &\\
				& $5^{2}S_{1/2}$ &8.679 &\\
				& $6^{2}S_{1/2}$ &9.044 &\\
				
				(S=3/2)& $1^{4}S_{3/2}$ &7.149 &7.130 &7.017 &7.187 &7.120\\
				& $2^{4}S_{3/2}$ &\textbf{7.497} & &7.571\\
				& $3^{4}S_{3/2}$ &7.829 & &7.985\\
				& $4^{4}S_{3/2}$ &8.148 &\\
				& $5^{4}S_{3/2}$ &8.455 &\\
				& $6^{4}S_{3/2}$ &8.752 &\\   
				
				\noalign{\smallskip}\hline\noalign{\smallskip}    
				
				(S=1/2)& $1^{2}P_{3/2}$ &7.261\\
				& $2^{2}P_{3/2}$ &\textbf{7.664}\\
				& $3^{2}P_{3/2}$ &8.047\\
				& $4^{2}P_{3/2}$ &8.412\\
				& $5^{2}P_{3/2}$ &8.762\\
				
				(S=3/2) & $1^{4}P_{5/2}$ &7.452\\
				& $2^{4}P_{5/2}$ &\textbf{7.655}\\
				& $3^{4}P_{5/2}$ &7.853\\
				& $4^{4}P_{5/2}$ &8.046\\
				& $5^{4}P_{5/2}$ &8.234\\
				
				\noalign{\smallskip}\hline\noalign{\smallskip}    
				
				(S=1/2) & $1^{2}D_{5/2}$ &7.480\\
				& $2^{2}D_{5/2}$ &\textbf{7.816}\\
				& $3^{2}D_{5/2}$ &8.135\\
				& $4^{2}D_{5/2}$ &8.444\\
				& $5^{2}D_{5/2}$ &8.741\\

				\br
			\end{tabular}\\
		\end{center}
	\end{table}
	\normalsize
	
	\subsection{Masses in the ($n,M^{2}$) plane.}
	So far we have calculated the ground and excited state masses of $\Xi_{cc}$, $\Xi_{bc}$, $\Omega_{cc}$, and $\Omega_{bc}$ baryons for natural and unnatural parity states in the $(J,M^{2})$ plane. In this section, we now expand our work by obtaining the radial and orbital excited state masses for these doubly heavy baryons in the $(n,M^{2})$ plane. Since the general equation for linear Regge trajectories in the $(n,M^{2})$ plane can be expressed as,
	
	\begin{equation}
		\label{eq32}
		n = \beta_{0} + \beta M^{2},
	\end{equation}
	where $n$ = 1, 2, 3.... is the principal quantum number, $\beta_{0}$, and $\beta$ represents the intercept and slope of the trajectories. The Regge slope ($\beta$) and Regge intercept ($\beta_{0}$) are same for the baryon multiplets lying on the single Regge line. We calculate $\beta$ and $\beta_{0}$ using relation (\ref{eq32}), and with the help of these parameters we estimated the excited state masses of these baryons lying on each Regge lines for natural and unnatural parity states. For $\Xi_{cc}^{++}$ baryon we can write from the slope equation, $\beta_{(S)} = 1/(M^{2}_{\Xi_{cc}^{++}(2S)}-M^{2}_{\Xi_{cc}^{++}(1S)})$, where $M_{\Xi^{++}_{cc}(1S)}$ = 3.579 GeV (calculated above) and taking $M_{\Xi^{++}_{cc}(2S)}$ = 3.920 GeV from \cite{Zalakcas_cc} for $1/2^{+}$ trajectory, we get $\beta_{(S)}$ = 0.30106 GeV$^{-2}$. From Eq. (\ref{eq32}) we can write, 
	
	\begin{eqnarray}
		\label{eq33}
		1 &= \beta_{0(S)} + \beta_{(S)} M^{2}_{\Xi^{++}_{cc}(1S)},\\ \nonumber
		2 &= \beta_{0(S)} + \beta_{(S)} M^{2}_{\Xi^{++}_{cc}(2S)},
	\end{eqnarray}
	using the above relations, we get $\beta_{0(S)}$ = -4.00918. With the help of $\beta_{(S)}$ and $\beta_{0(S)}$, we calculate the masses of the excited $\Xi^{++}_{cc}$ baryon for $n$ = 3, 4, 5... Similarly, we can express these relations for $P$ and $D$-wave as,
	\begin{eqnarray}
		\label{eq34}
		1 &= \beta_{0(P)} + \beta_{(P)} M^{2}_{\Xi^{++}_{cc}(1P)},\\ \nonumber
		2 &= \beta_{0(P)} + \beta_{(P)} M^{2}_{\Xi^{++}_{cc}(2P)},\\ \nonumber
		1 &= \beta_{0(D)} + \beta_{(D)} M^{2}_{\Xi^{++}_{cc}(1D)},\\ \nonumber
		2 &= \beta_{0(D)} + \beta_{(D)} M^{2}_{\Xi^{++}_{cc}(2D)},
	\end{eqnarray}
	Hence using the above equations, the masses for radial and orbital excited states from $1^{2}S_{\frac{1}{2}}-6^{2}S_{\frac{1}{2}}, 1^{4}S_{\frac{3}{2}}-6^{4}S_{\frac{3}{2}}, 1^{2}P_{\frac{3}{2}}-5^{2}P_{\frac{3}{2}}, 1^{4}P_{\frac{5}{2}}-5^{4}P_{\frac{5}{2}}$, and $1^{2}D_{\frac{5}{2}}-5^{2}D_{\frac{5}{2}}$   are estimated.
	In the same manner, we estimated the radial and orbital excited states of other doubly heavy  baryons for natural and unnatural parity states. The calculated results are summarized in tables \ref{table5}-\ref{table8}.
	
	\section{Results and Discussion}
	In the present work, we studied the mass spectrum of doubly heavy $\Xi$ and $\Omega$ baryons with two heavy quarks and one light quark ($ncc, scc, ncb,$ and $scb$) within the framework of Regge phenomenology. With the assumption of linear Regge trajectories, the relations between slope ratios, intercepts, and baryon masses are extracted and with the help of these relations, the ground as well as excited state masses of these baryons are estimated in both the ($J,M^{2}$) and ($n,M^{2}$) planes successfully. Our predicted masses for $\Xi_{cc}$, $\Xi_{bc}$, $\Omega_{cc}$, and $\Omega_{bc}$ baryons for natural and unnatural parity states in the ($J,M^{2}$) and ($n,M^{2}$) planes are summarized in tables \ref{table1}-\ref{table4} and \ref{table5}-\ref{table8} respectively. 
	
	Since $\Xi_{cc}^{++}$ is the only doubly heavy baryon observed experimentally assigned with spin-parity $J^{P}=\frac{1}{2}^{+}$ having mass 3.621 GeV in PDG \cite{PDG}. Our calculated ground state $\Xi_{cc}^{++}$ having mass 3.579 GeV is close to the experimental result with a mass difference of 42 MeV and it is also in good agreement with other theoretical prediction \cite{Ebert2002,Wei2015,Rosner2014,Alexandrou}. Our estimated masses 3.729 GeV and 3.726 GeV for $\Xi_{cc}^{++}$ and $\Xi_{cc}^{+}$ respectively for $J^{P}=\frac{3}{2}^{+}$ is close to the result of Ref. \cite{Ebert2002} with a mass difference of only 2 MeV and also in good agreement with other theoretical predictions \cite{Yoshida2015,Roberts2008,Wei2015,Rosner2014} with a mass difference in the range of 25-40 MeV. Further we compared our calculated results for radially and orbitally excited state masses of $\Xi_{cc}^{++}$  and $\Xi_{cc}^{+}$ baryons  with other theoretical outcomes as well. The predictions of Ref. \cite{Yoshida2015,Ebert2002} are in good agreement with our estimated masses in the ($J,M^{2}$) and ($n,M^{2}$) with a mass difference of few MeV (see tables \ref{table1} and \ref{table5}).  A recent search for the $\Xi_{bc}^{0}$ baryon by the LHCb Collaboration has been done in which no significant signal is found in the invariant mass range from 6.7-7.2 GeV/c$^{2}$. In this work, our estimated ground state mass 6.906 GeV lies within this range, which can be useful for future experimental searches. Also the obtained ground state ($1^{2}S_{\frac{1}{2}}$) mass is very close to the predictions of Refs. \cite{Rosner2014,Giannuzzi2009} with a difference of 2-8 MeV and also consistent with the results of \cite{Mathur N2018,Ebert2002,Z.S. Brown2014} with a mass difference of 27-40 MeV. Similarly, for $1^{4}S_{\frac{3}{2}}$ state our estimated mass is in good agreement with the results of Refs. \cite{Roberts2008,B. Eakins2012,Mathur N2018} having mass difference of 30-45 MeV. Further we compared the excited state masses for $\Xi_{bc}^{0}$ and $\Xi_{bc}^{+}$ baryons with various theoretical models. Since very few results are available and they are consistent with our predictions (see Tables \ref{table3} and \ref{table7}). 
	
	No experimental states have been observed in the case of doubly heavy $\Omega_{cc}$ and $\Omega_{bc}$ baryons yet. Many authors have made predictions about the masses of these baryons using various theoretical models so far. Our estimated results for ground and excited state masses of $\Omega_{cc}$ baryon in the ($J,M^{2}$) and ($n,M^{2}$) planes are summarized in tables \ref{table2} and \ref{table6} respectively. The computed ground state mass 3.719 GeV with $J^{P}=\frac{1}{2}^{+}$ is compared with the results of other theoretical models and it is found to be close to the predictions of Refs. \cite{Ebert2002,S. Migura2006} and similarly the obtained mass 3.847 GeV having $J^{P}=\frac{3}{2}^{+}$ is well matched with the outcomes of Refs. \cite{ZalakOmega_cc,Roberts2008,Ebert2002,Wei2015} having mass difference in the  range of 25-40 MeV. Further we compared our calculated excited state masses in the ($J,M^{2}$) and ($n,M^{2}$) planes and they are consistent with the results of Refs. \cite{Wei2015,Ebert2002,Roberts2008}. In the case of $\Omega_{bc}$, most of the authors have calculated the ground state masses only. Our predicted mass 7.035 GeV with $J^{P}=\frac{1}{2}^{+}$ is in good agreement with the estimated masses of the Refs. \cite{Mathur N2018,Giannuzzi2009,Roncaglia1995} with a mass difference of range 10-40 MeV.  Similarly the calculated mass 7.149 GeV with $J^{P}=\frac{3}{2}^{+}$  is close to the results of other theoretical  predictions \cite{ZalakOmega_cc,Ebert2002,Roberts2008,Roncaglia1995} with a mass difference of 29-38 MeV (see table \ref{table4}). The obtained excited state masses in the ($J,M^{2}$) plane for natural and unnatural parity states are compared with the results of shah \textit{et. al} \cite{ZalakOmega_cc} and we observed that our estimated masses for $1^{2}P_{\frac{3}{2}}$, $1^{2}D_{\frac{5}{2}}$, and $1^{4}P_{\frac{5}{2}}$ shows $\approx$ 60-100 MeV mass difference. Also, the radial and orbital excited state masses in the ($n,M^{2}$) plane are shown in table \ref{table8}. Only \cite{Giannuzzi2009} has calculated the radial excited states up to $3S$ and our obtained results for $2S$ and $3S$ are consistent with the prediction of \cite{Giannuzzi2009}. 
	
	\section{Conclusion}
	The Regge phenomenology has been employed successfully in obtaining the ground as well as excited state masses of doubly heavy $\Xi_{cc}$, $\Xi_{bc}$, $\Omega_{cc}$, and $\Omega_{bc}$ baryons in both the ($J,M^{2}$) and ($n,M^{2}$) planes. 
	In the present work we confirmed the spin parity of $\Xi_{cc}^{++}$  baryon with $J^{P}=\frac{1}{2}^{+}$, the only experimentally observed doubly heavy baryon.  Also the obtained mass 6.906 GeV for $\Xi_{bc}^{0}$ having $J^{P}=\frac{1}{2}^{+}$ lies within the invariant mass range 6.7-7.2 GeV/c$^{2}$ which is presented in a recent study on search of  $\Xi_{bc}^{0}$ baryon \cite{LHCbcas_bc} and our obtained result could definitely provide useful information in future experimental searches. The doubly heavy $\Omega$ baryons are yet to be observed by experiments. In a recent study of first search of $\Omega_{bc}$ baryon in the mass range of 6.7-7.3 GeV/c$^{2}$ presented by LHCb Collaboration, in which no signal is found within this mass range. The obtained mass value of $\Omega_{bc}^{0}$ baryon with $J^{P}=\frac{1}{2}^{+}$ as 7.035 GeV lies in this mass range which can also provide valuable contributions in future experimental searches. Also most of the theoretical studies predict the ground state masses of $\Xi_{bc}^{0}$ and $\Omega_{bc}^{0}$ baryons in between the mass range of 6.7-7.3 GeV \cite{Mathur N2018,Azizi2019,Zalakcas_cc,ZalakOmega_cc,Ebert2002,Roberts2008,Rosner2014,Roncaglia1995,Albertus}. Hence our mass value predictions for ground and excited states could be useful in future experimental studies at LHCb, CMS, Belle II, SELEX etc. to identify these baryonic states.
		
		In the present work we have calculated the natural and unnatural parity states in both the $(J,M^{2})$ and $(n,M^{2})$ plane. The model's reliability is very much depend upon the availability of experimental data that we have taken as a input.

	\section{Acknowledgments}
	The authors are thankful to the organisers of 10$^{th}$ international conference on new frontiers in physics (ICNFP 2021) for providing the opportunity to present our work.

\end{document}